\documentclass[conference]{IEEEtran}

\usepackage{cite}
\usepackage[pdftex]{graphicx}
\usepackage{amsmath}
\usepackage{algorithmic}
\usepackage{array}
\usepackage[caption=false,font=footnotesize]{subfig}
\usepackage{url}
\usepackage{booktabs}
\usepackage{multirow}
\usepackage{xspace}
\newif\ifcommentson
\commentsontrue 

\newcommand{\myi}{\textit{(i)}\xspace}
\newcommand{\myii}{\textit{(ii)}\xspace}
\newcommand{\myiii}{\textit{(iii)}\xspace}

\usepackage{footmisc}

\newif\ifanonymous
\begin{document}

\title{Self-Balancing Semi-Hierarchical PCNs for CBDCs}
\author{
    \IEEEauthorblockN{
        Marco Benedetti, 
        Francesco De Sclavis,
        Marco Favorito,
        Giuseppe Galano,\\
        Sara Giammusso,
        Antonio Muci,
        Matteo Nardelli$^*$
    }
    \medskip
    \IEEEauthorblockA{Technical Report CFC.CRYPTO.CS/2024/1\\
    Applied Research Team (ART) - IT Department - Bank of Italy$^\dag$}
}

\maketitle

\renewcommand{\thefootnote}{\fnsymbol{footnote}}
\footnotetext[1]{E-mail address of the authors are in the form [firstname].[lastname]@bancaditalia.it, except for  giuseppe.galano2@bancaditalia.it.}
\footnotetext[2]{The views expressed in this paper are those of the authors and do not necessarily reflect those of the Bank of Italy.}
\renewcommand*{\thefootnote}{\arabic{footnote}}

\begin{abstract}
We introduce a family of  PCNs (Payment Channel Networks) characterized by a semi-hierarchical topology and a custom set of channel rebalancing strategies. This family exhibits two interesting benefits, if used as a platform for large-scale, instant, retail payment systems, such as CBDCs:
Technically, the solution offers state-of-the-art guarantees of fault-tolerance and integrity, while providing a latency and throughput comparable to centralized systems;
from a business perspective, the solution perfectly suits the 3-tier architecture of the current banking ecosystem (central banks / commercial banks / retail users), assigning a pivotal and peculiar role to the members of each tier.
Furthermore, the cryptographic privacy of payments for retail users---typical of PCNs such as the public Lightning Network---is largely (possibly fully) retained.
We study the system by simulating a scaled-down version of a hypothetical European CBDC, exploring the trade-offs among liquidity locked by market operators, payment success rate, throughput, latency, and load on the underpinning blockchain.
% Results are encouraging.
\end{abstract}

\section{Introduction}
\label{sec:intro_MB}

Retail instant payment systems---such as credit/debit card circuits or Central Bank Digital Currencies (CBDCs)---manage a very high load of transactions, in the order of $10^4$--$10^5$ TPS (see Sect.~\ref{sec:background_throughput_MB}).  
Centralized systems are often deployed to ingest and settle all such transactions timely.

Performance is not all though: 
In the special case of CBDCs, further business and technical requirements are put into place\footnote{For the case of the digital euro, see e.g.~\cite{bce:de_market_research,bce:stocktake_de}}, such as the possibility to set a cap on the amount of liquidity users can amass (wallet cap), strong privacy guarantees for all payment data, small or no fees for citizens (``\textit{free for basic use}''), and the possibility to be usable by unbanked people.

Finally, retail instant payment systems have to be embedded into---and play nice with---the pre-defined, multi-tier structure of the existing banking and payment infrastructure:
a pyramid that sees the Central Bank(s) at the top, a set of authorized intermediaries (such as commercial banks and similar financial institutions) in the middle, and retail users (such as citizens and merchants) at the bottom---providing each actor with clear and strong incentives to adopt and use the system itself\footnote{ For an introduction to this three-tier structure, see\cite{mishkin2007economics}.}.

We ask: Can a blockchain-based payment system address all these concerns and requirements while bringing to the table tangible benefits in comparison to a centralized solution?

In this work, we move the first steps towards tentatively answering such a question in the positive (for a CBDC).

First essential point: Most blockchains are well-known for their limited throughput, so on-chain settlement of retail payments is very difficult to achieve. We forgo such a perspective entirely and embrace the off-ledger paradigm, whereby scalability is achieved by an additional ``payment channel network'', or PCN (called $2^{nd}$ layer) built on top of the actual blockchain (see Sect.~\ref{sec:PCN_MB}).
In the wild, $2^{nd}$ layer networks---e.g., the Lightning Network~\cite{Poon:2016}---evolve freely, as peer-to-peer systems: They are unstructured and adopt best-effort payment routing strategies.
Unfortunately, their anatomy \myi is not coherent with the 3-tier structure of the monetary system, \myii is oblivious to most of the business requirements of a CBDC, and \myiii provides a level of scalability that---albeit significant w.r.t. the underpinning blockchains---is still insufficient for a large-scale retail instant payment system.

%in which our payment infrastructure is to be embedded, nor capable of sustaining the required throughput.

%But, in a CBDC-like scenario, any $2^{nd}$ layer network 
%must be embedded into the pre-defined, multi-tier structure of the existing banking and payment infrastructure:
%
%The resulting architecture is neither coherent with the structure of the financial market in which our payment system is to be embedded, nor capable of sustaining the required throughput. 
%

%Our work is motivated precisely by the mismatch between the shape and performance of freely formed PCNs on one side, and the anatomy and requirements of CBDCs on the other side (See Section~\ref{sec:motivation_MB}).

We address these issues by devising a family of PCNs with a partially constrained topology, called ``Semi-Hierarchical Payment Channel Networks'', or SH-PCNs (Sect.~\ref{sec:SH_PCN_MB}), endowed with a set of custom balancing strategies (Sect.~\ref{sec:auto-rebalance-MB}). SH-PCNs are designed to play nice with the financial ecosystem and to exhibit performances akin to a retail payment system.

Assuming SH-PCNs allow us to achieve the required throughput and latency with sufficient reliability (a question we ponder in Sect.~\ref{sec:PCN-simulator}), we can shift our attention to the benefits of using a distributed, blockchain-based, 2-layer solution:
Blockchains introduce tools capable of increasing integrity, availability, fault-tolerance, and verifiability of monetary exchanges; at the $2^{nd}$ (retail) layer, they \myi accommodate, by design, levels of privacy similar to that of cash; \myii naturally handle wallet caps; \myiii give the central bank (CB) and the market control on the transaction fees for retail users.

We do not model a full-scale SH-PCN system just yet, but a scaled-down version. This reduced model is still capable of exhibiting the key dynamics we want to investigate, such as the trade-offs among the liquidity locked in channels, the success rate of payments, their latency, and the transactional demand on the underlying $1^{st}$-layer blockchain (see Sect.~\ref{sec:related}).

Our preliminary results suggest that the performance of SH-PCN is satisfactory for retail payments; that SH-PCNs map naturally onto the banking/payment ecosystem; and that the most typical CBDC-specific requirements are naturally met.

In the following sections, we provide a description of the solution that carefully combines the technical, PCN-oriented angle with the business, CBDC-oriented perspective.

%In particular, we focus on the network re-balancing strategies that allow an SH-PCN to work reliably under a sustained load of a given magnitude.

%
\begin{figure*}
    \centering
    \includegraphics[width=0.95\textwidth]{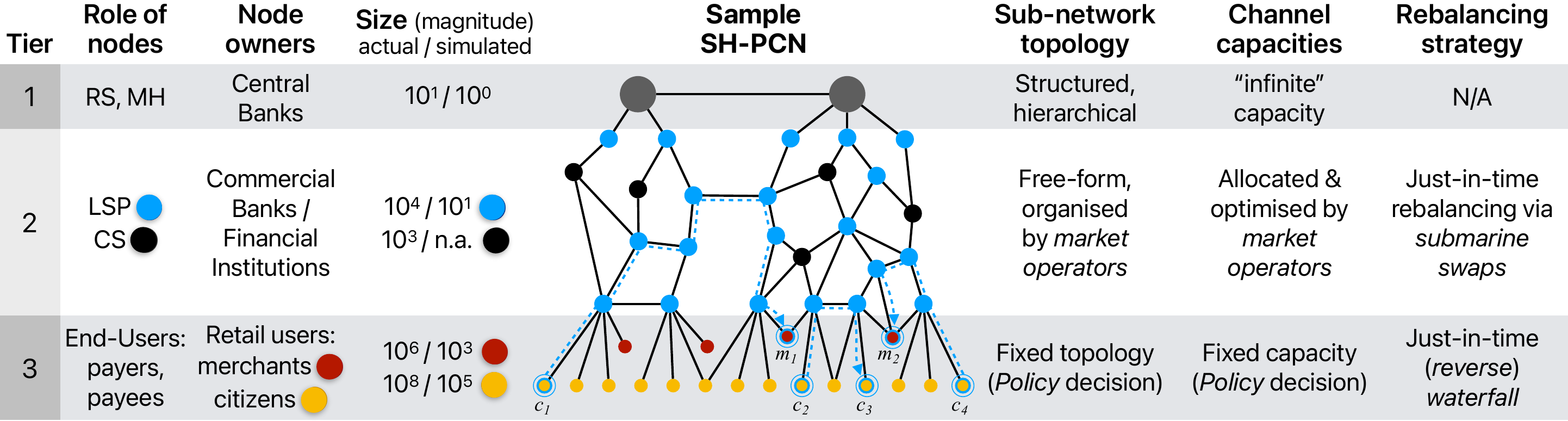}
    \caption{Structure and main features of a Semi-Hierarchical Payment Channel Network (SH-PCN), plus three sample payment routes: One \textit{Point-of-Sale} payment from retail user $c_4$ (a citizen) to merchant $m_2$; one \textit{Peer-to-Peer} transaction from $c_2$ to $c_3$; one \textit{eCommerce} route from  $c_1$ to a foreign merchant $m_1$. As usual with source-based onion-routed protocols, no actor (other than the payer and payee) has  full information about any of these payments.
    %each LSP along these paths only knows about its surroundings (previous and subsequent nodes in the route);   
    }
    \label{fig:SH-PCN}
\end{figure*}

\section{The Payment Channel Network (PCN) model}
\label{sec:PCN_MB}

In a PCN, two nodes create a bilateral \emph{payment channel} by ``pre-funding'' it, i.e., by locking some amount of liquidity, called \emph{channel capacity}, into a 2-of-2 multisig UTXO, using a single \emph{on-chain} funding transaction; 
the sum of nodes' balance in the channel can never exceed the payment channel capacity.

After the channel is created (i.e., the funding transaction is confirmed), the nodes can exchange multiple, instant, off-chain payments, which update the balances of nodes and eliminate the need to constantly execute on-chain transactions.

The safety of payments within the channel is ensured through the previously described pre-funding and a cryptographic mechanism called \emph{channel revocation}: If a cheating attempt by one party is detected, the other is entitled to claim the entire channel capacity both as compensation and as a penalty for the attempted cheating.
For a comprehensive description of how PCNs work and how they relate to the underpinning blockchain, we refer the reader to~\cite{Poon:2016,Gangwal23}.

The largest PCN in existence is the public Lightning Network (LN). Recent studies (e.g.,~\cite{Rohrer:19,Seres:20}) show that the topology of the LN exhibits the characteristics of a small-world and scale-free network.
As shown in~\cite{Avarikioti:18}, such scale-free topologies are sub-optimal in terms of locked liquidity. 
And while an alternative star-like network configuration would help decrease liquidity inefficiencies, it would also reduce fault-tolerance, security, and privacy of payments~\cite{Rohrer:19}.

%LSPs function as the payment equivalent of Internet Service Provider (ISP) in the communication domain.
%CSs can be seen as the payment equivalent of Cloud Service Provider in communication domain.

%
\section {Structure of a Semi-Hierarchical PCN}
\label{sec:SH_PCN_MB}

The current monetary system---which is the target deployment environment for our solution---is based on a \emph{three-tier banking system}. 
The pyramid sees the CB at the top, a set of authorized intermediaries (such as commercial banks and similar financial institutions) in the middle, and retail users (such as citizens and merchants) at the bottom. For an introduction to this three-tier architecture, see\cite{mishkin2007economics}.

These different (business) roles will be mapped onto various different (technical) roles in the PCN subclass we study, and the hierarchical anatomy of the banking system will be reflected into the (semi) hierarchical topology of the network.

%So, while in the public LN all nodes are (in principle) peers to one another, the SH-PCN we define in this work has (i) different classes of nodes and (ii) a partially constrained topology, as in Figure~\ref{fig:auto-repl:submarine-swap-MB}.

%In particular, Central Banks would play a key role in our system, by supplying the (central bank) liquidity necessary to set up payment channels---and by running the underlying blockchain where such ``wholesale CBDC'' would be created and managed, whereas traditional financial institutions (e.g., banks) would constitute a semi-structured network of payment channels among them---and channels towards retail users---whereby they act as \myi payment routers, and \myii trusted and regulated crossing points between central bank retail money and commercial bank money (e.g., deposits).

%
In SH-PCNs, we make a distinction among four classes of technical nodes, managed by four types of business actors, organized into \textit{3 tiers}: 
\textbf{Monetary Hubs} (MHs) at the top,
\textbf{Lightning Service Providers} (LSPs) and \textbf{Custodian Services} (CSs) in the middle,
and \textbf{End-Users} (EUs) at the bottom, 
as represented in Fig.~\ref{fig:auto-repl:submarine-swap-MB}. %In detail:
\begin{itemize}
\item\textbf{Monetary Hubs}. Tier-1 nodes typically managed by a CB (or a set of CBs in the same monetary area): At the on-chain layer, they are fully responsible for the operation of the ledger where transactions to open/close channels are stored and where (CB) liquidity is locked. At the off-ledger layer, they operate nodes with considerable liquidity, which are largely (or fully) interconnected, and which act as \emph{Routing Service Providers} (RSPs) for LSPs in Tier-2. Thus, MHs act as business-to-business entities and do not interact with EUs;
\item \textbf{Lightning Service Providers}. Tier-2 nodes acting as the core, distributed ``service providers'' of the system. LSPs freely open channels towards each other, forming a ``small world''-like network, ensuring that fault tolerance and the other benefits of distributed networks are achieved. Disconnected LSP sub-networks, if they exist, are linked via RSP(s). From a business perspective, this role is assigned to banks and other financial institutions, which are properly supervised and own large amounts of liquidity.
LSPs offer their service to EUs by opening channels with them, thus providing them with connectivity and reachability within the payment network.
\item \textbf{Custodian Services}. In addition to off-chain CB liquidity in the SH-PCN, EUs generally own one or more accounts at Tier-2 CSs, such as banks or exchanges, which allow EUs to deposit/withdraw retail CBDC to/from their personal accounts\footnote{These account-based ledgers are managed by a (centralized) system entirely decoupled from the PCN, under the responsibility of the CS.}, which amass ``commercial bank money''. The CSs are connected to the LSPs network. The EU does not necessarily have direct channels opened with CSs, as their monetary interactions happen either out-of-band (see Sect.~\ref{sec:waterfall} and~\ref{sec:reverse_waterfall}) or through LSP intermediaries.
\item \textbf{End-users}. Tier-3 nodes managed by ``end-users'' (i.e., citizens and merchants): 
% They are the ultimate beneficiaries of the payment service. 
They access the network connecting to one or more LSPs in a \emph{non-custodial} way,
% i.e., they retain control over their private keys and route calculations, 
thereby \textit{preserving the privacy of their payments}.
Crucially (from a business perspective) we have that \myi EUs do not open channels with each other\footnote{This is not a technical limitation, but a policy option: If end users were allowed to access the Layer 1 blockchain to freely open channels among them, the level of privacy of payments would quickly approach full anonimity.}, and \myii EU channels---with a known capacity---are only opened with one or more authorized LSP(s): We can place a cap on the maximum amount of retail CBDC present in the user wallet, without sacrificing privacy. 
\end{itemize}
So, Tier-1 is the domain of CBs; Tier-2 is where market operators act; EUs live at Tier-3, in a hierarchy where MHs only connect to LSPs and EUs only connect to LSPs and CSs.
The number of EUs is much larger than the number of LSPs/CSs, which in turn are many more than the MHs. 
%
%See Figure~\ref{}.

%
\section{Automated Channel Rebalancing Techniques}
\label{sec:auto-rebalance-MB}

Absent any dynamics in network topology (see Sect.~\ref{sec:simplifying_assumptions_MB}), the performance (success rate, throughput, latency) of SH-PCNs subject to most types of sustained loads of payments among end users degrades over time: Channels become more and more unbalanced, hence unable to participate into the routing of an increasing percentage of payments.

To avoid such progressive occlusion, we define 3 rebalancing techniques, aimed at improving channel lifetimes while minimizing locked liquidity.
We make the hypotheses that \myi LSPs and routing nodes try to keep their channels balanced by proactively issuing trustless ``\emph{submarine swaps}'' between on-chain and off-chain liquidity; \myii end users 
% (i.e., retail users and merchants)
rebalance their channels in real-time, when needed, by withdrawing funds from---or depositing funds to---their trusted CS(s)\footnote{These regulated entities are the points where AML/CFT activities can be performed, possibly issuing Suspect Transactions Reports (STRs) to the CB. }.
%(e.g., a bank or an exchange).
%
% \hlr{1.3 col}
\vspace{-2pt}
\subsection{Submarine swaps between two LSP/MH nodes}
A \emph{submarine swap} is a transaction that exchanges a given amount of some on-chain digital asset with the same amount of the off-chain form of the same asset (e.g., $x$ bitcoin are moved from $A$ to $B$ on the blockchain and ``simultaneously'' $x$ bitcoin go from $B$ to $A$ in the Lightning Network).

Submarine swaps can be trustless: At no point during the process either party has access to the other party's fund, and no third party takes temporary custody of the asset. The swaps are atomic, with a binary outcome: Either the two parties successfully exchange their assets, or the swap fails.

\begin{figure}[h!]
    \centering
    \includegraphics[width=0.9\columnwidth]{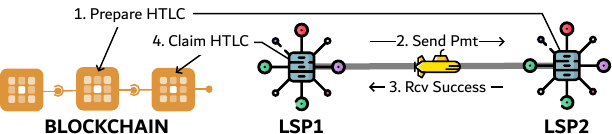}
    \caption{Atomic submarine swap functionality}
    \label{fig:auto-repl:submarine-swap-MB}
\end{figure}
To implement the trustless requirement of a submarine swap, both the on-chain and the off-chain transactions rely on Hash-Time Locked Contracts (HTLCs). 
Fig.~\ref{fig:auto-repl:submarine-swap-MB} represents two LSPs that share an unbalanced channel, with almost all liquidity held at the LSP1 end. LSP1 wants to rebalance the channel by transferring assets on the payment channels with LSP2, who in turn should move the same amount of assets on-chain to LSP1. 
They proceed as follows.
LSP2, sender of the on-chain transaction, generates a preimage, 
% which is essentially a random 32-byte secret
whose hash serves as the foundation for constructing both the on-chain and off-chain HTLCs, and moves funds to the on-chain HTLC (step 1). 
LSP2 also generates an off-chain invoice using the same hash and asks LSP1, sender of the off-chain transaction, to pay the invoice. 
After the chain confirms the locking of funds in the on-chain HTLC, LSP1 can safely pay the invoice (step 2).
To claim the off-chain funds, LSP2 has to reveal the preimage to LSP1 (step 3), who in turn can use it to complete the process, by claiming the funds from the on-chain HTLC (step 4).
% 
%HTLCs ensures the safety of the submarine swap.
%
At any time, if any of the two parts misbehaves, the other can get its funds thanks to the unlocking conditions of the HTLCs.

Due to the reliance on on-chain transactions, a submarine swap takes time (at least the confirmation time of the HTLC in step 1) and possibly requires a fee (depending on the policy enacted by the MH). Moreover, the transactional capacity of the underlying blockchain limits the number of submarine swaps per unit of time that can be performed by the SH-PCN.

\subsection[Waterfall rebalance]{Waterfall\footnote{The terms ``\textit{waterfall}'' / ``\textit{reverse waterfall}''  are lifted from~\cite{bce:stocktake_de}.
%
%They designate processes for which we provide a PCN-oriented, fully compliant, implementation.
We present a PCN-oriented, fully compliant implementation of the processes they designate.} rebalance}
\label{sec:waterfall}
%
%The waterfall and reverse waterfall mechanisms are inspired by and address the requirements reported in~\cite{bce:stocktake_de}
%
The \emph{waterfall} mechanism allows end-users---in particular those having high inbound traffic (e.g., merchants)---to always be able to get paid, even if the amount $P$ to be received raises the user balance $B$ above the channel capacity limit $C$ (i.e., the wallet cap). This is achieved by automatically depositing the amount $D = \max{(B+P-C, L_D)}$, where $L_D$ is a minimum amount that the user is willing to deposit, to a linked CS account. 
\begin{figure}[h!]
    \centering
    \includegraphics[width=0.9\columnwidth]{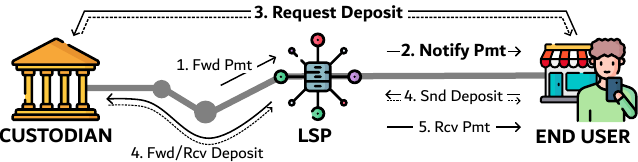}
    \caption{Waterfall functionality implementation in a PCN}
    \label{fig:auto-repl:waterfall}
\end{figure}
Fig. \ref{fig:auto-repl:waterfall} illustrates the waterfall functionality. 
When the LSP receives a payment to forward (step 1), but the end user's channel does not have enough outbound liquidity, the LSP \emph{notifies} the user about the incoming payment (step 2), and delays the payment forwarding until the expiration of a timeout. The user requests a real-time deposit to their CS (step 3), which in turn sends the deposit via the same LSP, thus rebalancing the channel (step 4). If the channel is successfully rebalanced within the timeout, the LSP forwards the payment to the user (step 5); otherwise, the payment fails. 
Messages exchanged in steps 2 and 3, in bold, are not part of the LN specification, but are specific to our protocol.
% between end user and respectively their custodian and their LSP that
%
From a business perspective, this mechanism implements a deposit of retail CBDC liquidity from the wallet of an EU into an account held by the same EU at some CS, in commercial bank money, triggered when the CBDC wallet is about to overflow the cap. 

\subsection{Reverse waterfall rebalance}
\label{sec:reverse_waterfall}
\begin{figure}[h!]
    \centering
    \includegraphics[width=0.9\columnwidth]{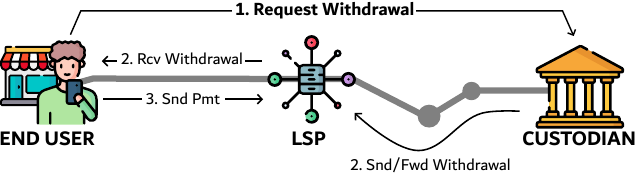}
    \caption{Reverse waterfall functionality implementation in a PCN}
    \label{fig:auto-repl:rev-waterfall}
\end{figure}

The \emph{reverse waterfall} functionality allows retail users---in particular those having high outbound traffic (e.g., citizens)---to automatically fund a payment channel before making transactions too large for the current state of that channel.
Fig. \ref{fig:auto-repl:rev-waterfall} illustrates the reverse waterfall functionality. 
If there are insufficient funds in the payment channel to cover the payment amount $P$, users could request a withdrawal (step 1), thus taking out some money from their CS account (step 2). The withdrawal amount is $W = \max{(L_W-B, P-B)}$, where $L_W$ represents a minimum amount that the user is willing to keep in its wallet for future use. Once the withdrawal has successfully transferred liquidity from the linked CS to the payment channel, the user can send the payment (step 3). 
The message exchanged in step 1 is not part of the LN specification, but is specific to our protocol.
From a business perspective, this mechanism implements a withdrawal of commercial bank money from the account of an EU at their CS, which is returned as CBDC liquidity put into the wallet of the same EU; this process is triggered when the CBDC wallet has insufficient funds to operate.

%The waterfall and reverse waterfall mechanisms are inspired by and address the requirements reported in~\cite{bce:stocktake_de}.
% \GG{ Vogliamo citare il doc BCE che descrive waterfall e rev waterfall da cui abbiamo preso nomi e requisiti? \url{https://www.ecb.europa.eu/paym/digital_euro/investigation/profuse/shared/files/dedocs/ecb.dedocs231018.en.pdf}} \MN{Ok}
% 
% \MN{Il concetto di submarine swap è noto nella letteratura PCN. Non lo è il concetto di waterfall e reverse waterfall. Non mi è chiaro se vogliamo presentarli come nostri (quindi in sez\ref{sec:auto-repl}) o introdurli come background.}

\section{The real world and our small-scale model}
\label{sec:real_world_and_model}

\subsection{Patterns of real-world retail payments} 
\label{sec:background_throughput_MB}

Retail payment systems manage a very high load of transactions. A commercial, global-scale example is VISA, which recently disclosed that its network can execute more than $65,000$ transactions per second at its peak\footnote{\url{https://usa.visa.com/solutions/crypto/deep-dive-on-solana.html}}. 

For a public retail payment system, such as a prospective CBDC for the euro area, we can refer to the ``\emph{Study on the payment attitudes of consumers in the euro area}'' (SPACE) by the ECB~\cite{bce:space}.
The study reports a survey among a randomly selected sample of the population across $19$~euro area countries. 
Participants were requested to document their \textit{point-of-sale (POS)}, \textit{peer-to-peer (P2P)}, and \textit{online} transactions in a one-day diary. 
The survey also includes {recurring payments} that are out of the scope of our investigation. 
The key results of SPACE 2022 are summarized in Table \ref{tab:ecb-space-table-MB}.
% \usepackage{multirow}

In addition, we consider the Annex 1 of the ECB's ``\emph{Market Research on Potential Technical Approaches for a Digital Euro}''~\cite{bce:de_market_research}, which states that, on average, individuals within the euro area engage in two financial transactions per day, encompassing various payment methods and interaction points.
The ``Large'' adoption case considered in this report assumes that 70\% of the eurozone population will use the CBDC for 35\% of all transactions.
This implies approximately 2,000 TPS on average.
%, assuming 70\% of the Eurozone population as a user base. 
Peak values may be larger by an order of magnitude.
% 2*344000000*0,70*0,35/(24*60*60) - scenario Large

These numbers---in the $10^3$--$10^5$ range---corroborate our working hypothesis about the need of an off-ledger layer to face retail payments. And, they provide us with a clear reference point to designing proportioned scaled-down models.

The aforementioned SPACE report, as seen in Table~\ref{tab:ecb-space-table-MB}, provides us with further interesting information about real-world payments, which we can use to assemble a believable small-scale model: the distribution of transaction amounts, the proportion among Point-of-Sale, eCommerce, and Peer-to-Peer payments, and the tendency of retail payments to be domestic (within the same country) rather than cross-border.

Finally, it is worth recalling that a payment is defined as ``instant'' in this context if it completes within 10 seconds~\cite{ecbinstant:17}.
\begin{table}
\caption{Statistics on non-recurring payments. ECB SPACE 2022.}
\label{tab:ecb-space-table-MB}
\setlength{\tabcolsep}{2.5pt}
\renewcommand{\arraystretch}{1.25}
\begin{tabular}{|l|c|c|c|c|c|c|c|c|}
\hline
\multicolumn{1}{|c|}{\multirow{2}{*}{Type}} &
  \multirow{2}{*}{\begin{tabular}[c]{@{}c@{}}N.\\ (\%)\end{tabular}} &
  \multicolumn{7}{c|}{Distribution (\%) of Range Amount (€)} \\ \cline{3-9} 
   &
   &
  $<5$ &
  $[5,10)$ &
  $[10,20)$ &
  $[20,30)$ &
  $[30,50)$ &
  $[50,100)$ &
  $>100$ \\ \hline
PoS &
  $80$ &
  $21$ &
  $17$ &
  $21$ &
  $13$ &
  $13$ &
  $10$ &
  $5$ \\ \hline
Online &
  $17$ &
  $10$ &
  $11$ &
  $20$ &
  $15$ &
  $17$ &
  $16$ &
  $11$ \\ \hline
P2P &
  $3$ &
  $(14)^*$ &
  $(11)^*$ &
  $(22)^*$ &
  $(16)^*$ &
  $(14)^*$ &
  $(11)^*$ &
  $(12)^*$ \\ \hline
\end{tabular}
\rule{0pt}{1em}    
\footnotesize{$^*$ Unfortunately, the frequencies of amount ranges in the P2P scenario are not available in SPACE; these values have been estimated by the authors.}\\
\end{table}

\subsection{Size of our small-scale model}
\label{sec:small_scale_model}
We study a network on a scale $\sim$1:1000 relative to the actual population of the euro area (344 million citizens).
Our network thus consists of 300k individuals (retail users) at Tier 3.
%, assuming that the observed statistical behaviors hold true also in a full scale networks.

We incorporate a merchant every 100 people, and a service provider every 10k end-users, in line with the 6.4k inhabitants per bank branch in the EU~\cite{EBF:22}, for a total of 3k merchants and 30 service providers in Tier 2.
We model 3 countries, representative of different scales found in the euro zone: a small country (such as Cyprus, with $\sim$1M citizens), one medium-size country (such as Finland, $\sim$6M), and a large country (such as Italy, $\sim$60M).
Each user and each intermediary is associated with one of these countries, and the distribution of nodes among countries is proportional to their population.

At Tier 1, each country has a (node representing its) national Central Bank (NCB). NCBs live in the same monetary area and are interconnected.
Cross-national payments are assumed to be a fraction of the national payments, namely 5\%.
Overall, there are $3$ countries (CBs), $3\cdot 10^1$ service providers (financial institutions), $3\cdot 10^3$ merchants, and $3\cdot 10^5$ citizens.

The number of payments per second is also scaled 1:1000, proportionally to the population: Given the average 2,000 TPS from Sect.~\ref{sec:background_throughput_MB}, we simulate 2 TPS for 24 hours, or $172,800$ payments, which move $9.75$M units of value. When we study a peak load, we move an order of magnitude up, to 20 TPS (for 12 hours), thus to $950,000$ payments ($53.2$M in value).

In our model, we scaled down the number of users, and this brought about a proportional reduction in both the number of payments per second they generate, and in the size of the routing network. We do \emph{not} scale quantities that do not vary with the size of the population: The cap on end-user wallets is taken as-is from what has been hypothesized in~\cite{bindseil2020tiered},~\cite{bindseil2020central} and~\cite{meller2023know}, i.e., €3000; the distribution of payment amounts is harmonized with the figures in Table~\ref{tab:ecb-space-table-MB}; the definition of ``instant'' payment still requires completion within 10s~\cite{ecbinstant:17}.

\subsection{Simplifying assumptions}
\label{sec:simplifying_assumptions_MB}
We adopt several simplifying assumptions to facilitate the initial study of our SH-PCN under high load. 
We assume that:
\begin{itemize}
    \item the topology of the SH-PCN is static, i.e., after the initial setup, no new channels are opened, nor are existing ones closed, during the time window of interest;
    \item each EU, whether a person or a merchant, has only a single channel, with a single LSP;        
    \item LSPs are connected to their reference MH/RSP in such a way that full connectivity is ensured;
    \item the LSP of an EU also acts as their reference, trusted CS (the roles of LSP and CS are collapsed on the same actor);    
    \item EUs, LSPs, RSPs, CSs and EUs are always online;
    % , while end users receiving payments can be temporarily offline;
    \item fees are not imposed to route payments; so the best payment route is the shortest path between the payer and the payee (and not the one occasioning the smallest fee);
    \item the blockchain introduces a constant delay (modeled on the speed of actual permissioned fault-tolerant systems, such as~\cite{Benedetti23}) when on-chain transactions are needed.
    % \item we simulate and study a scaled-down version of the system, according to the following Sect.~\ref{sec:small_scale_model}.
\end{itemize}

\section{Modeling and simulating SH-PCNs}
In order to study how SH-PCNs behave if used to support a retail payment system akin to the one described in Sect.~\ref{sec:real_world_and_model}, given the rebalancing strategies from Sect.~\ref{sec:auto-rebalance-MB}, we model the entire system, and we simulate its execution over a full business day, under different load and liquidity conditions.

Our architecture is composed of 4 logical sub-models: 

\begin{itemize}

\item[\textbf{A}.] The \textbf{SH-PCN topology model} and generator (Sect.~\ref{sec:model_topology}), tasked with creating realistic SH-PCNs; the network it creates is given as input to:

\item[\textbf{B}.] The \textbf{SH-PCN runtime model} and simulator (Sect.~\ref{sec:PCN-simulator}), which simulates the working of actual PCNs to great detail; this simulator is set in motion by an incoming stream of payment requests, generated by:

\item[\textbf{C}.] The \textbf{load model} and generator for retail payments (Sect.~\ref{sec:model_payment}), which outputs 24 hours worth of simulated payments, according to Sect.~\ref{sec:small_scale_model}; as the PCN evolves under the payment load, it is analyzed and updated by:

\item[\textbf{D}.] The \textbf{PCN realtime rebalancer} (Sect.~\ref{sec:PCN-rebalancing}), which monitors the PCN and applies the strategies  from Sect.~\ref{sec:auto-rebalance-MB} to keep the system running at full success rate.
\end{itemize}

Sub-models (A) and (C) are inspired to the structure, throughput, and functioning of actual European payment systems/patterns, although we operate at a reduced scale (Sect.~\ref{sec:small_scale_model}) and under simplifying assumptions (Sect.~\ref{sec:simplifying_assumptions_MB}).

\subsection{SH-PCN topology model and generator}
\label{sec:model_topology}
% chatgpt generated

We designed a generator of random SH-PCN instances, such as the one in Figure~\ref{fig:SH-PCN}. The generator takes as an input a list of parameters---such as the number of entities to place in each tier, the constraints on their inter-connections, the capacity of channels among different actors, etc---and generates one random SH-PCN instance with the given features (in a human-friendly JSON format).
While the generator is capable of constructing 1:1 networks, using the parameters from Section~\ref{sec:background_throughput_MB}, we use the small scale variant from Section~\ref{sec:small_scale_model} to generate the SH-PCN we experiment on in  Section~\ref{sec:experiments}.

Our SH-PCN random model is a composition of various random graph models, each representing the connections between different categories of nodes and within the same category:
\begin{enumerate}
    \item Central banks are interconnected in a clique formation, with each link having a capacity of €500 million.
    
    \item The interconnections between LSPs are modeled using a Watts-Strogatz graph model \cite{watts1998collective}, using an average degree $k=4$ and a rewiring probability $p=0.1$; the capacities associated with channels in this subnetwork are not fixed: A range of different values for such capacities is explored in the experimental section.
    
    \item The links between CBs and LSPs are generated by dividing the LSPs into distinct subsets, each corresponding to a CB/country. Subsequently, a channel is established from each CB to every LSP in its respective subset. The subset sizes is log-normally distributed (with $\mu=0$ and $\sigma=1$); the capacity of such CB-LSP channels is not fixed: The simulation explores a range of possibilities.

    \item The creation of channels between LSPs and end users, encompassing both citizens and merchants, is approached similarly to the CB-LSP connections, albeit with \textit{fixed} capacities, which represent the cap on users' wallets (CBDC requirement).
    Specifically, channels connecting LSPs with citizens have a capacity of €3000. In contrast, channels linking LSPs with merchants vary based on the merchant's size: small (S), medium (M), and large (L) merchants are assigned channels with capacities of €5k, €50k, and €500k, respectively.
\end{enumerate}
Overall, here is a summary of the submodels in a SH-PCN:
\begin{table}[h]
\centering
%\caption{Random Graph Models of the Subnetworks in our SH-PCNs}
\label{tab:my_label}
\renewcommand{\arraystretch}{1.2}
\begin{tabular}{|c|c|c|}
\hline
\textbf{Connection Type}  & \textbf{Random Graph Model} & \textbf{Capacity} \\ \hline
Among CBs & Clique & €500M \\ \hline

CBs to LSPs & Log-normal ($\mu=0$, $\sigma=1$) & several \\ \hline

Among LSPs & Watts-Strogatz ($k=4$, $p=0.1$) & several \\ \hline

LSPs to Citizens & Log-normal ($\mu=0$, $\sigma=1$) & €3000 \\ \hline

LSPs to Merchants & \multirow{2}{*}{Log-normal($\mu=0$, $\sigma=1$)}   &  \multirow{2}{*}{€5$\cdot10^{3 / 4 / 5}$} \\

(S / M / L) &  &  \\ \hline
\end{tabular}
\end{table}

\subsection{SH-PCN runtime model and simulator}
\label{sec:PCN-simulator}
Several PCN simulators exist in the literature (see e.g. \cite{Beres:19,DiStasi:18,Rebello:22,Yu:18,Conoscenti:21cloth}). No one dominates the others along all dimensions, as  different authors rely on the simulation of specific features of the PCN protocol, but not others.
For our purposes, we selected CLoTH~\cite{Conoscenti:21cloth}, which stands out as an open-source sequential Discrete Event Simulation (DES) engine that reproduces the code of a mainstream lightning node implementation\footnote{\url{https://github.com/lightningnetwork/lnd}}, particularly focusing on routing and on the HTLC mechanics.

CLoTH features a straightforward simulation run-time and is used in several papers (see e.g.,~\cite{Asgari:22,Davis:22}). 
However, it lacks support for parallel processing. We tested its serial engine and realized it does not perform adequately when very large networks and large payment volumes are to be analyzed. 

Therefore, we developed a CLoTH-like simulator within ROSS~\cite{Ross:github_docs}, a Parallel DES. ROSS operates on a distributed-memory architecture, and can run on multiprocessor systems.
We exploit its native advanced features, e.g., parallel execution, optimistic event scheduling, and load distribution. 

\begin{figure}[t]
    \centering
    \includegraphics[width=1\columnwidth]{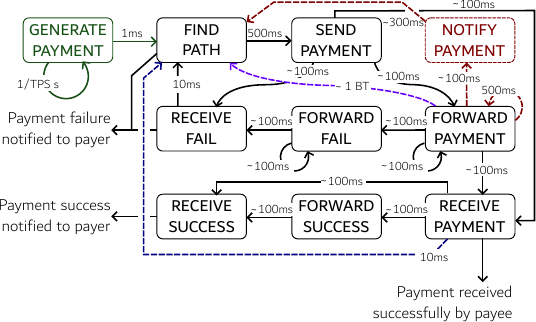}
    \caption{Event diagram of our ROSS-based, CLoTH-like PCN simulator.}
    \label{fig:sim:event-state-diagram-MB}
\end{figure}

\vspace{4pt}\textbf{CLoTH on ROSS.}
Within ROSS, we implement and reuse the same events and states as CLoTH: Fig.~\ref{fig:sim:event-state-diagram-MB} contains the event diagram of our PCN simulator. The elements and edges in black are part of the core CLoTH simulation. Elements in color represent extensions to CLoTH that we introduce: One for dynamic load generation---in green; two for the waterfall and reverse waterfall functionalities---respectively in red and blue; one for the submarine swap functionality---purple.

For each event, we implemented all the relevant ROSS callbacks: the event handler, the reverse event handler used in optimistic simulations, and the commit event handler that is used to permanently update the state of LPs. These handlers call the appropriate (and properly wrapped) CLoTH code.

\vspace{4pt}\textbf{Executing a payment.}
When the payment request is received, a new \texttt{FIND PATH} event is scheduled for the payment process (or a withdrawal if the payer does not have enough liquidity to execute the payment, see Sect.~\ref{sec:PCN-rebalancing}). The \texttt{FIND PATH} computes the path between the sender and the receiver of the payment. If a path cannot be found, the payment fails. Otherwise, the payment follows the same life-cycle as in CLoTH, going through the sequence \texttt{SEND PAYMENT}, \texttt{FORWARD PAYMENT} (if the payment traverses more than one edge), \texttt{RECEIVE PAYMENT}, \texttt{FORWARD SUCCESS} (if any) and \texttt{RECEIVE SUCCESS} for a successful payment. Conversely, a non-successful payment goes through \texttt{FORWARD FAIL} (if any) and \texttt{RECEIVE FAIL}. A payment may fail due to an insufficient balance to forward it along the path selected by the sender. Once a \texttt{RECEIVE FAIL} notifies the failure to the sender, the payment can be re-tried by scheduling a new \texttt{FIND PATH} event. This loop is allotted 10 seconds, after which the process is aborted and the payment fails definitely.

\vspace{4pt}\textbf{From SH-PCN nodes to ROSS processes.}
ROSS uses two structures to represent processing units:
\begin{itemize}
    \item A \emph{Logical Process} (LP), which serves as the fundamental computation unit. LPs manage their own event queue, receive messages, process the events they contain, and optionally send additional messages to other LPs.
    \item A \emph{Physical Process} (PE), which are the actual processes of the distributed MPI architecture, to which LPs run on. PEs orchestrate the parallel simulation and manage the communication with other PEs for the LPs they host.
\end{itemize}

In our simulation we have an LP for each node in the SH-PCN (all included: Tier 1, 2, and 3 nodes are simulated). We include a single auxiliary LP for each PE that orchestrates the payment generation for its own LPs (see Sect.~\ref{sec:model_payment}).

\vspace{4pt}\textbf{Parallelizing the simulation.}
In ROSS, simulations can be executed sequentially or in parallel, using either a \textit{conservative} or an \textit{optimistic} approach. We employ an optimistic simulation, whereby ROSS activates a synchronization mechanism called ``Time Warp'', which uses a detection-and-recovery protocol to synchronize the computation: Any time an LP detects out-of-order events, it rolls those events back and re-executes them.

Load distribution is crucial in a distributed-memory PDES. Co-locating chatty LPs on the same PE significantly improves performance, as messages can be exchanged in memory instead of relying on inter-process communication. 
For our PCN simulation we devise a domain-specific mapping strategy. 

First, we construct a weighted graph of nodes, where edges are labeled with an estimated communication frequency. Then, in order to achieve a balanced distribution of citizens, merchants and intermediaries among partitions, we assign weights to each node. Following the approach outlined in~\cite{gill2021high}, we partition the resulting graph using METIS~\cite{metis}, which minimizes the sum of the weights of the edge traversing partitions and equalizes the total node weights among partitions.

\subsection{Model and generator for retail payments}
\label{sec:model_payment}
The payment generator schedules a \texttt{GENERATE PAYMENT} event (green in Fig.~\ref{fig:sim:event-state-diagram-MB}) with a frequency that depends on a target global payment rate.
As detailed in Sect.~\ref{sec:small_scale_model}, the global rate in the small scale model is between 2 and 20 PPS (Payments Per Second).
Such global rate is apportioned to PEs, in such a way that each PE generates a local load that is proportional to the number of retail users co-located on the PE itself.
Then, the payment generator creates a single random payment between two users that are randomly selected according to the SPACE model described in Sect.~\ref{sec:real_world_and_model}.

We assume that geographically close users exhibit a greater tendency to transact with each other, hence users predominantly conduct domestic transactions within their own country, while cross-border ones occur infrequently.

The load generator selects the payment sender first, which is always a retail user. Then, it selects a random payment scenario according to the SPACE distribution of payment contexts (POS, eCommerce, P2P) and to the static cross-border payment probability.
Depending on the selected payment scenario, the generator selects a random receiver among merchants (for POS and eCommerce payments) or retail users (for P2P payments).
Finally, the payment generator chooses the transaction amount according to the SPACE conditional probability distribution of amounts (see Table~\ref{tab:ecb-space-table-MB}). 

\subsection{PCN real-time rebalancer}
\label{sec:PCN-rebalancing}

The rebalancer applies the strategies described in Sect.~\ref{sec:auto-rebalance-MB} to the payment system simulated via our custom blending of CLoTH and ROSS, described in Sect.~\ref{sec:PCN-simulator}.
The technical implementation of the strategies is as follows.

\vspace{4pt}\textbf{Waterfall.} 
Whenever a payment that is about to be received by an EU would result in the channel balance exceeding its cap, the payment is notified to the receiver using the \texttt{NOTIFY PAYMENT} event, and the processing of the \texttt{FORWARD PAYMENT} event is delayed. Upon receiving the notification, the payee initiates another transaction, i.e., a \emph{deposit}, to transfer the excess liquidity to a CS. The deposit process is triggered by the \texttt{FIND PATH} event, using a delay that takes into account the roundtrip time for the deposit request. Then, \texttt{FIND PATH} computes the (best) path between the sender of the deposit (receiver of the payment) and their CS. If the deposit is completed before the original payment expires, the \texttt{FORWARD PAYMENT} can be completed and the payment can be received successfully. Otherwise a \texttt{FORWARD FAIL} event is generated. See the red state/links in Fig.~\ref{fig:sim:event-state-diagram-MB}.

\vspace{4pt}\textbf{Reverse Waterfall.} 
When a payment is generated, and the sender has an insufficient balance, the simulation initiates a new transaction, i.e., a \emph{withdrawal}, from the sender's associated CS to the sender itself, using a \texttt{FIND PATH} event. The original payment is placed in a queue of payments awaiting withdrawals, that is managed by the sender. Once the withdrawal is successfully received (\texttt{RECEIVE PAYMENT} event), the sender of the original payment triggers it creating the respective \texttt{FIND PATH} event.
See the blue link in Fig.~\ref{fig:sim:event-state-diagram-MB}.

\vspace{4pt}\textbf{Submarine Swap.}
A submarine swap is initiated by either an LSP or an RSP in response to a \texttt{FORWARD PAYMENT} event when a payment about to be forwarded would result in the channel balance exceeding a predefined \emph{unbalancedness} threshold (e.g., $ 80\% $ of its capacity). The simulation then has the LSP generate a \texttt{FIND PATH} event towards its channel counterparty, with a delay of 1 block time, as outlined in the purple line of Fig.~\ref{fig:auto-repl:submarine-swap-MB}. The off-chain payment simulates the submarine leg of the swap, depicted in steps 2 and 3 of Fig.~\ref{fig:auto-repl:submarine-swap-MB}, while the block time delay of the event simulates the on-chain HTLC preparation of step~1.
See the purple link in Fig.~\ref{fig:sim:event-state-diagram-MB}.

\section{Experimental Evaluation}
\label{sec:experiments}
Payments in PCNs can fail for a number of reasons, the main one being the insufficient capacity of some link along the route chosen by the payer/source. 
When a route fails, another one is tried, until success---or until some failure criterion is met, e.g., a timeout expires.
As the load on the system increases, the probability of failure is expected to increase, both because channels may become unbalanced and because there is an increasing resource contention from multiple in-flight concurrent payments with partially overlapping routes.

In this context---fully captured by our simulation---we show (a) under which conditions our rebalancing strategies are able to keep the payment success rate at 100\% under a constant load; (b) that payments remain ``instant'' in all cases; (c) how the system reacts to a sudden 10x increase in the number of payments per seconds; (d) how the trade-off between liquidity locked in channels and their rebalancing plays out. 

\subsection{Payment Success Rate}
\label{sec:exp-PSR}
\begin{figure}[t]
    \centering
    \includegraphics[width=1.0\columnwidth]{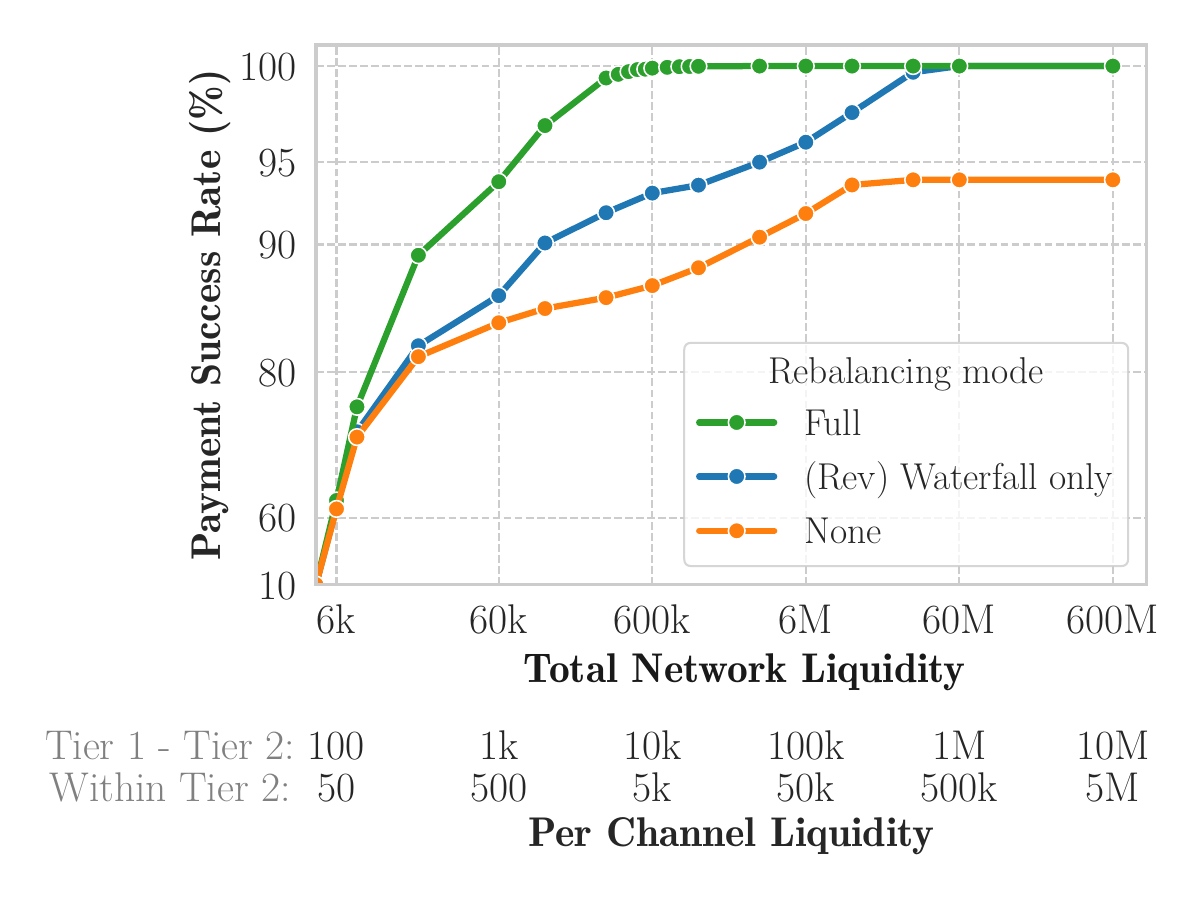}
    \caption{Percentage of payments routed instantly by the SH-PCN ($y$ axis) as a function of the liquidity in the network/channels ($x$ axis). Log-log scale.}
    \label{fig:experiment-PSR}
\end{figure}
We present our SH-PCN with a constant transaction load as per Sect.~\ref{sec:model_payment}, exerted for 24 hours (simulation time).

%our SH-PCN can achieve a close-to-100\% payment success rate within 10 seconds (definition of ``instant payment''), under a constant transaction load, exerted for 24 hours.

Figure~\ref{fig:experiment-PSR} shows the success rate of payments as a function of the \emph{total amount of liquidity} Tier 1 and 2 operators are willing to lock into the system. By ``total amount'', we mean the sum of the capacity of all the channels among Tier 2 operators, plus the channels between Tier 1 and Tier 2\footnote{The other capacities are not accounted for, because: Capacities from Tier 2 to Tier 3 (i.e., wallet caps) are set by policy and are not subject to adjustments for performance reasons; channels within Tier 1 nodes are practically infinite (managed by CBs); channels among Tier 3 entities do not exist in our setting.}.

The success rate is non-zero ($\sim$16\%) even for zero total liquidity, because there is a fraction of payments that happens between retail customers of the same Tier 2 entity, which can be routed successfully without traversing internal Tier 2 links. 

As the liquidity of the network grows, so does the percentage of payments successfully regulated within ``instant'' timespans.
In our small-scale model, a fully self-rebalancing network (green line) achieves a 100\% success rate after approximately 600k units of total network liquidity.
If we disable the rebalancing tool internal to Tier 2 (submarine swaps) and keep active the Tier 2-3 rebalancing (waterfall, direct and inverse), we are in a scenario where payments can always be started and received, but they can possibly fail to be routed due to inner channels becoming permanently (almost) unbalanced.
In this case (blue line), to reach the same success rate, 
we need to inject hundreds of times more liquidity into the network ($\sim$60M units).
Finally, the removal of all rebalancing mechanisms results---as expected---in a failure of the system to even approach a 100\% success rate (orange line).

\subsection{Instant-ness of Payments}
\label{sec:exp-instant}
\begin{figure}[t]
    \centering
    \includegraphics[width=1.0\columnwidth]{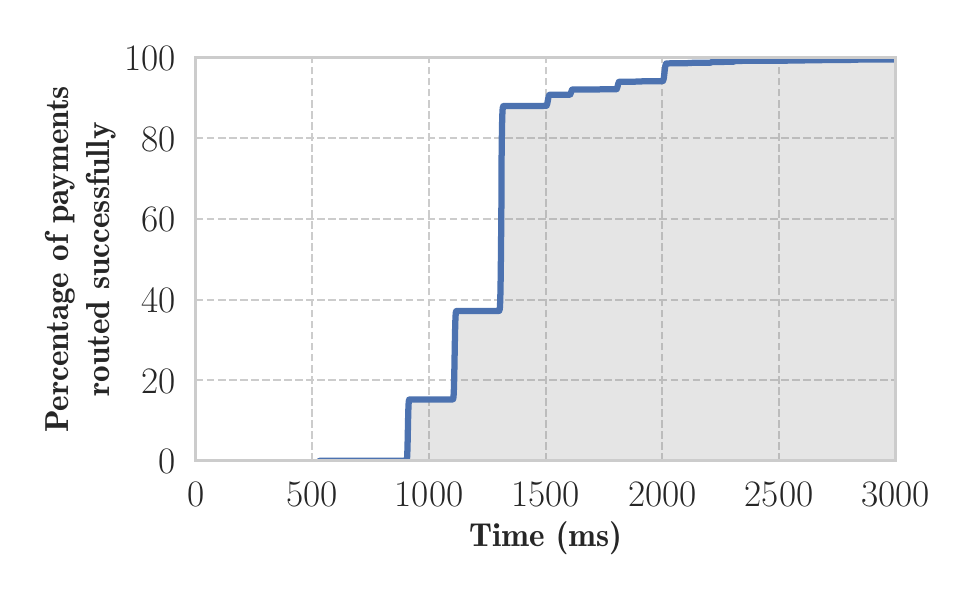}
    \caption{Time-to-completion of payments: Cumulative distribution.}
    \label{fig:experiment-instant}
\end{figure}
Figure~\ref{fig:experiment-instant} reports the cumulative distribution of completion times for retail transactions, averaged over a full day of simulated payments, i.e.,
the curve represents the amount of payments that succeed within a given amount of time.
Any capacity beyond the 400k threshold identified in Sect.~\ref{sec:exp-PSR} produces essentially the same results.
The bumpiness of the curve hinges on our simulation being time discrete, with events introducing delays as in Fig.~\ref{fig:sim:event-state-diagram-MB}.
% only happening once every $\sim$100ms.
%
We observe how all payments complete within at most 3 seconds, well under the 10 second cutoff that identifies instant payments in our context.
%on the first try (first route attempted) in a given amount of time. The green curve represents what happens with repeated runs, as per the PCN protocol, within the 10s cutoff.

\subsection{Adaptive Rebalancing}
\label{sec:exp-adaptive}
\begin{figure}[b]
    \centering
    \includegraphics[width=1.0\columnwidth]{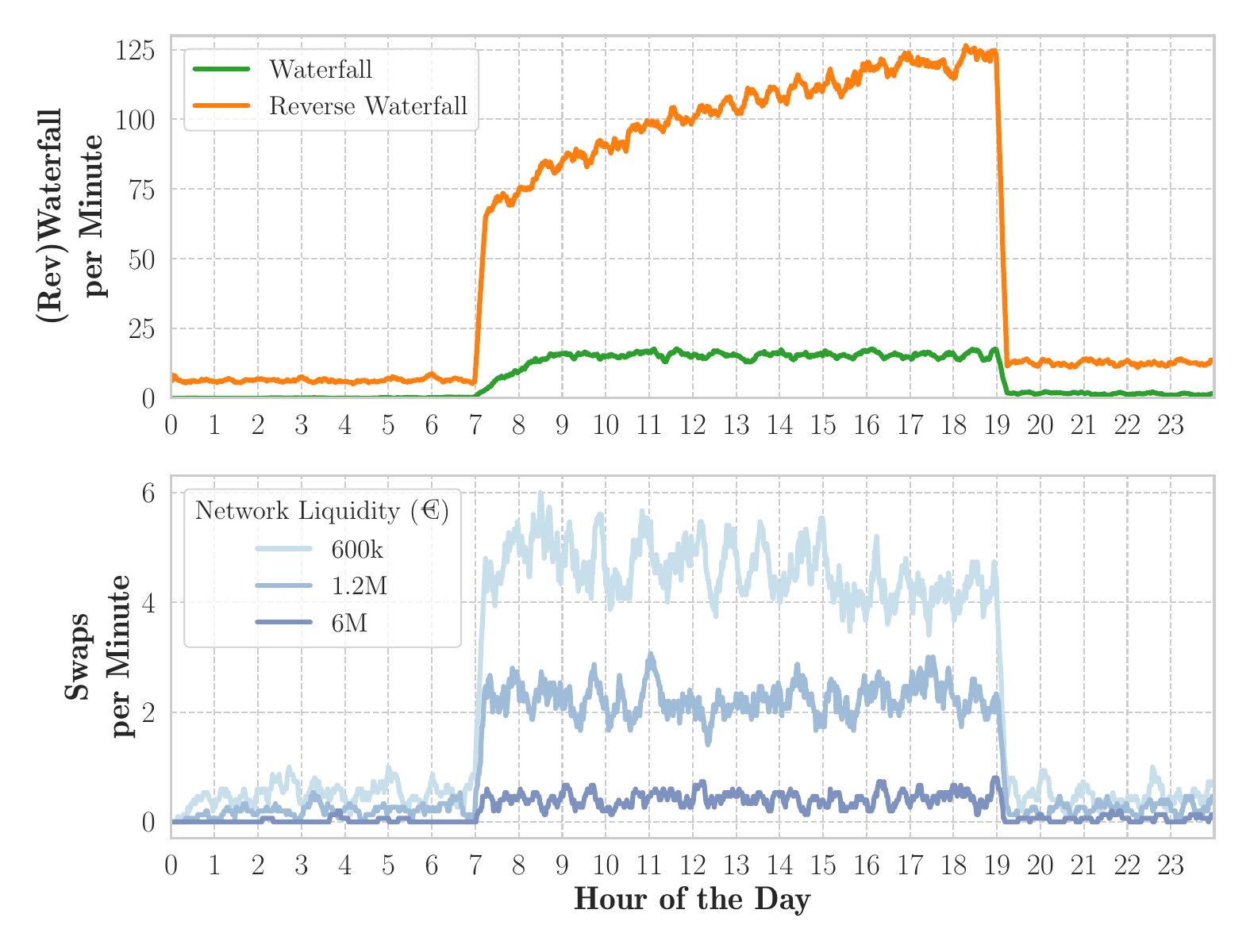}
    \caption{In the top chart we plot the number of (reverse) waterfall events per minute during a 12h stress test (from 7am to 7pm) under peak load.
    The (much smaller) number of submarine swaps per minute requested by the Layer 2 payment network to Layer 1 in order to keep itself balanced is reported in the bottom chart, for three different levels of total routing liquidity.}
    \label{fig:experiment-adaptive2}
\end{figure}
All the experiments shown so far have been performed under a constant load of transactions, equal to the average expected throughput for a payment system at our scale, according to the payment model described in Sect.~\ref{sec:small_scale_model}.
Actual payment systems must be able to sustain peaks that may increase the volume of payments by an order of magnitude over the nominal average throughput (the ``Christmas day'' effect). To show how our SH-PCN responds to one such occurrence, we simulate a day where from 7am to 7pm the load on the system suddenly increases by an order of magnitude.

Figure~\ref{fig:experiment-adaptive2} shows the system response: All the rebalancing mechanisms start working harder to maintain the entire payment network sufficiently well balanced, even under pressure.

As expected, the number of submarine swaps, for a given routing liquidity grows substantially under pressure. And, channels of larger capacity require fewer swaps, according to an approximately linear inverse relation.

In addition, Fig.~\ref{fig:experiment-adaptive2} shows that---as expected---the number of (reverse) waterfall events is independent of the network liquidity and of the number of submarine swaps (it only depends on the load). Also, it is much higher than the number submarine swaps, because the capacity/cap for retail wallets is much smaller than the average channel size.

The payment success rate (not shown in the chart) stays at 100\% at all times during the peak window; our simulated small-scale system is capable of maintaining  satisfactory throughput, latency, and performance, also under load.

\subsection{The cost of keeping a channel well balanced}

%Given these results, we continue our evaluation with a fully rebalanced SH-PCN having a total routing liquidity of 100k.

%Figure~\ref{} reports the distribution of completion times for retail transactions, averaged over a full day of simulated payments. The blue curve represents the amount of payments that succeed on the first try (first route attempted) in a given amount of time. The green curve represents what happens with repeated runs, as per the PCN protocol, within the 10s cutoff.

We have seen the inverse correlation between channel capacity and swaps per unit of time, at the channel level. At system level, the inverse correlation is between the total routing liquidity and the transactional demand on Layer 1.

\begin{figure}[t]
    \centering
    \includegraphics[width=1.0\columnwidth]{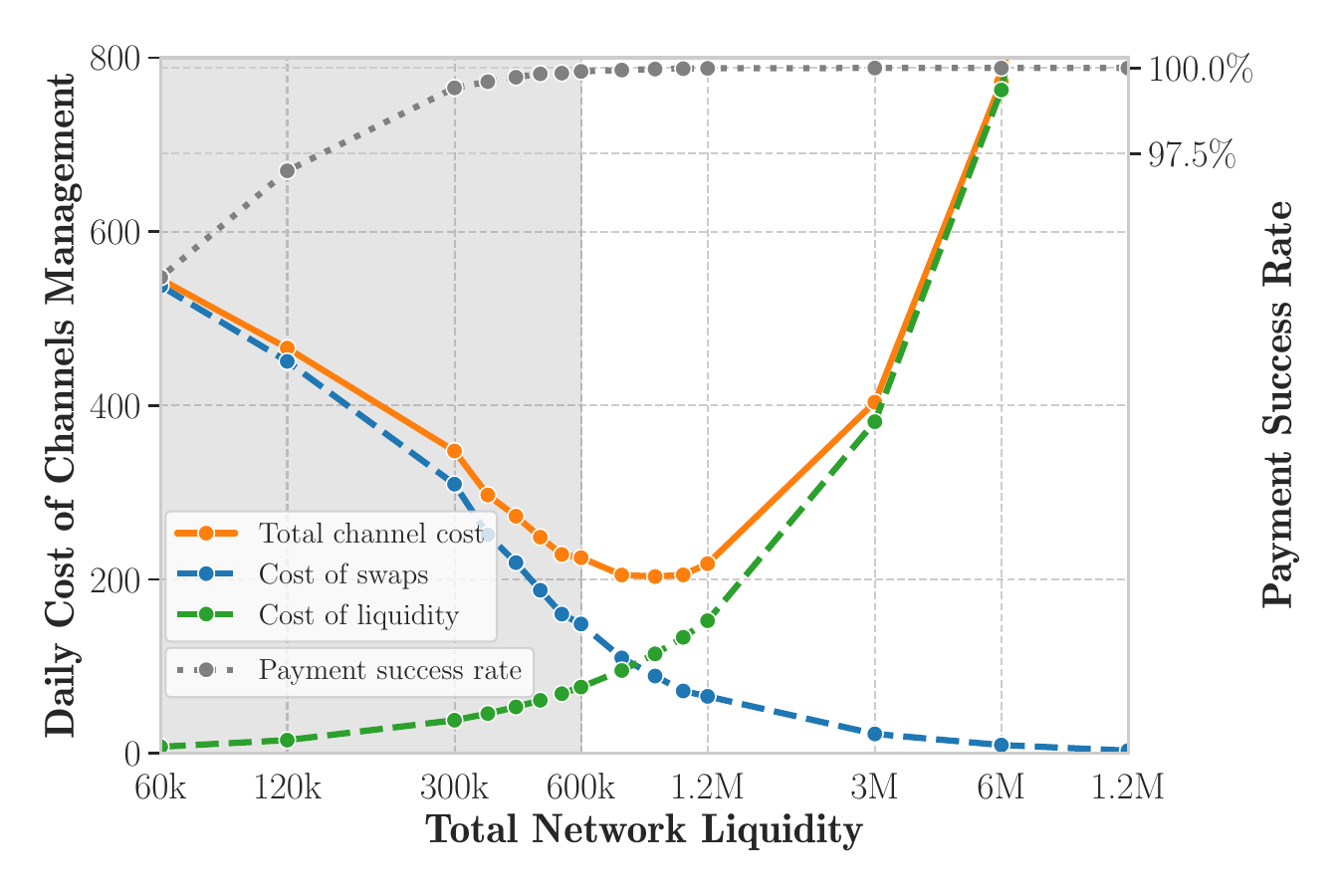}
    % in alternativa: 
    % \includegraphics[width=1.0\columnwidth]{img/CoL_v8_swap0.10x2-liquidity4.75year.pdf}
    \caption[Daily cost of locked liquidity and of submarine swaps for different network capacities.]{Daily cost of locked liquidity and of submarine swaps for different network capacities. The sum of these components (total cost of channel management) reaches a minimum at $\sim800$k. We assume the CB lends money\footnotemark at $4.75\%$, and charges $0.1$ monetary units for each Layer 1 transaction.}
    \label{fig:tradeoff}
\end{figure}

This technical trade-off engenders a related economical trade-off: Submarine swaps have an economic cost in the form of Layer-1 transaction fees, which are set by Tier 1 actors (i.e., CBs); at the same time, the cost for Tier 2 market operators to keep central bank liquidity locked into channels is again decided by Tier 1 actors, i.e., CBs. The resulting economic dynamics are out of scope for the present work, and we will not investigate in any detail the policy options for the CB.
We just show that, once the cost parameters have been set by Tier 1, the market can optimize (statically, or even dynamically under the current transaction load) the network liquidity at Tier 2.
In particular, the sum of (a) the cost of liquidity---that increases monotonically with the capacity of channels, plus (b) the cost of actuating submarine swaps---that increases monotonically with the number of swaps executed, is a convex function with a global minimum point.
If the minimum is to the left of the point where the system reaches 100\% success rate (gray area in Fig.~\ref{fig:tradeoff}), it means swaps are so cheap compared to borrowing liquidity that the actual lower bound for the network capacity is given by the minimum liquidity necessary to sustain a 100\% success rate (approx. 600k in our example).
Conversely, if the minimum is to the right of the point where submarine swaps are no longer needed (because the initial channel capacity suffices to support all payments in the 24 hours), it means liquidity is so cheap compared to executing swaps, that the optimal value for network liquidity is the minimum value sufficient to process all payments without swaps ($\sim30$M in our example).
\footnotetext{This seemengly arbitrary number is the annual Marginal Lending Facility (MLF) of the European Central Bank (ECB) as of Sept. 20, 2023.}
%
%which can leave the value of locked liquidity regulated by the general interest rate, or even decide to reward specifically the wholesale liquidity used by the market to provide a retail payment service to citizens.
%
The only remaining, interesting case is represented in Fig.~\ref{fig:tradeoff}: Here, the optimal amount of liquidity to invest by Tier 2 actors for the purpose of routing retail payments is $800k$, which falls in the non-trivial range $[600k;30M]$. This is where our small-scale SH-PCN requires the smallest running costs.

\section{Related Work}
\label{sec:related}

% \hlr{0,7--1 col}

% \subsection{Topology Definition}
% \hlr{goals: easy of management? success rate, privacy, transaction fee}
% 
The PCN topology affects the ability to successfully route payments, network resilience (against both attacks and random failures), and the privacy of payments.
The unstructured topology of LN led to a small-world and scale-free network, with a few highly-connected nodes that route most of the payments (e.g.,~\cite{Rohrer:19,Seres:20}).
Avarikioti et al.~\cite{Avarikioti:18} show that this topology is not optimal: when all transactions are known a-priori, a star topology can minimize the locked capital and maximize profits.
Such a central node reduces the average path length and thereby the routing failures due to unavailable balance, at the price of weakening fault-tolerance, security, and privacy 
(e.g.,~\cite{Kappos:21,Sguanci:22}).
% (e.g.,~\cite{Erdin:21,Kappos:21,Sguanci:22}).
% 
Rohrer et al.~\cite{Rohrer:19} show how the LN can be subjected to channel exhaustion or node isolation attacks.
Conversely, we resort to (semi) hierarchical topologies, which promise to combine the benefits of both previous topologies.
% (see Sect.~\ref{sec:SH_PCN_MB}).

% \subsubsection{Capital Assignment}
A key problem is to determine the initial capacity of channels: The bigger the channel capacity, the higher the number of routed transactions, the higher the liquidity costs (e.g., interest charges, opportunity costs).
% \subsubsection{Channel Replenishing and Rebalancing} 
% % 
To route a transaction, a channel moves liquidity, getting progressively imbalanced over time (thus reducing the success probability of routing).
We believe that adopting rebalancing techniques is a key pillar for highly loaded networks, which otherwise would need large initial investments to open channels.
Shabgahi et al.~\cite{Shabgahi:23} propose a model to predict the expected time for a channel to get unbalanced, considering the channel centrality and its initial balance.
To extend the channels lifespan, they can be either replenished with on-chain transactions (e.g., \cite{Sivaraman:18}) or rebalanced in-place 
(e.g., \cite{Awathare:21,Avarikioti:21,Hong22}). 
% (e.g., \cite{Pickhardt:20,Khalil:17,Awathare:21,Avarikioti:21,Hong22}). 
%
Centralized approaches to rebalance channels requires nodes to disclose their individual contribution in the channel, thus violating privacy (e.g., \cite{Khalil:17}). Avarikioti et al.~\cite{Avarikioti:21} enhance the protocol using privacy preserving techniques. 
Li et al.~\cite{Li:20} propose to divide time in epochs and to use a randomly selected committee of nodes to estimate the channels' capacity needed to route the expected payments in the next epoch. This approach incurs in a large number of on-ledger transactions.
% to re-full the channels' capacity. 
% 
Awathare et al.~\cite{Awathare:21} propose a decentralized protocol that leverages network cycles to perform circular payments and rebalance capacity. 
Focusing on relay nodes, Papadis et al.~\cite{Papadis23} use reinforcement learning to proactively perform submarine swaps aiming to maximize profit from fees.
In this paper, we study the impact of the initial channel capacity as well as the benefits of three different approaches for automatic rebelancing of channels. Although we rely on reactive solutions, we study their feasibility by considering constraints imposed by the layer-1 blockchain.
% the possibility of rebalancing channels' capacity, therefore we compute the initial channel assignment by considering the trade-off between liquidity cost, rebalancing cost, and success rate of transactions.}
% 
Interestingly, Miller et al.~\cite{Miller:19} propose an alternative to LN, enabling incremental deposits and withdrawal through extended off-chain contracts (i.e., with no impact on-ledger). Nonetheless, this approach is not compatible with LN. 
We postpone as a future work the compatibility analysis of our automated rebalancing techniques with this approach.   

% \subsection{Simulating LNs}
% 
So far, a rather large number of LN simulators exist (e.g., \cite{Beres:19,DiStasi:18,Rebello:22,Yu:18,Conoscenti:21cloth}), even though a clear comparison among them is lacking. 
% A large number of open-source simulators appears not to be actively maintained by the initial developers or by the community (e.g., \cite{Beres:19,DiStasi:18}).
% 
% One of the LN developer open-sourced a simulator aimed to study the ability to scale of the LN\footnote{\url{https://github.com/rustyrussell/million-channels-project}}; it only considers the LN gossiping algorithms and its message protocol.
% 
% Beres et al.~\cite{Beres:19} develop a simulator specifically focused on fee and profitabilty, name LNTrafficSimulator\footnote{\url{https://github.com/ferencberes/LNTrafficSimulator}}, to empirically study LN's transaction fees and privacy provisions. 
% Rebello et al.~\cite{Rebello:22} introduce an open-source PCN simulator that, building on the OMNET++ framework, implements the official Lightning message protocol. It consists of four main modules, which deal with generating the topology, generating workload, simulating the network, and visualizing/storing simulation results. The authors use this simulator to evaluate different routing algorithms. 
% Likewise, Yu et al.\cite{Yu:18} rely on a PCN simulator based on the ns-3 network simulator, aimed to evaluate a novel payment routing mechanism (i.e., CoinExpress). Its source code is not publicly available. 
% 
CLoTH~\cite{Conoscenti:21cloth} is an open-source simulator of LND, the most popular LN implementation, which has been used in different recent research works (e.g.,~\cite{Asgari:22,Davis:22}).
% ,Dasaklis:23
% 
Aiming to analyze very large-scale networks, we integrate CLoTH functionalities in ROSS, a general-purpose PDES (see Sect.~\ref{sec:PCN-simulator}).
% \hlb{To the best of our knowledge, none of the previous works analyzed the PCN locked liquidity--payments trade-off.}

\section{Conclusion and Future Work}
\label{sec:conclusion}

We introduced a class of self-rebalancing  semi-hierarchical PCNs, and showed experimentally---on a small-scale model based on the payment patterns and structure of the euro area---that they seem to (a) ensure good performances (in terms of latency, throughput, and success rate of payments) and (b) suit the existing 3-tier banking/payment architecture naturally.

As a future work, we will move towards removing one by one the simplifying assumptions listed in Sect.~\ref{sec:simplifying_assumptions_MB}, up to the point of presenting and studying a simulation at scale $1:1$, exhibiting a rich set of behaviors and dynamics. 

Finally, we will study and characterize the strenght of the privacy enjoyed by the retail payer and payee in this system, and compare it to public PCNs such as the Lightning Network.

% \bibliographystyle{IEEEtran}
% \bibliography{biblio}

% \IEEEtriggeratref{8}
\bibliographystyle{IEEEtran}
\bibliography{brain24}

\end{document}